\documentclass[preliminary,copyright,creativecommons,sharealike,noncommercial]{eptcs}
\providecommand{\event}{F-IDE 2018} 

\usepackage{wrapfig}

\newcommand{\WP}{\textrm{WP}}
\newcommand{\kw}[1]{\texttt{#1}}

\usepackage{tikz}
\usetikzlibrary{decorations.pathreplacing}
\usetikzlibrary{shapes,arrows,shadows}
\tikzstyle{bloc} = [rectangle, draw, fill=green!30,
    text centered, rounded corners]
\tikzstyle{tool} = [rectangle, draw, fill=red!30,
    text centered]
\tikzstyle{blocl} = [rectangle, draw, color=black!50, fill=green!10,
    text centered, rounded corners]
\tikzstyle{tooll} = [rectangle, draw, color=black!50, fill=red!10,
    text centered]
\tikzstyle{line} = [draw, -triangle 45]
\tikzstyle{linel} = [draw, color=black!50,-triangle 45]
\tikzstyle{lineb} = [line width=1.5pt, >=latex, ->,
shorten <=2pt, shorten >=2pt]

\usepackage{why3lang}
\lstdefinestyle{myAda}{language=Ada, keywordstyle=\color{blue}, morekeywords={some}}

\newcommand{\wa}[1]{\lstinline[style=myAda,basicstyle=\ttfamily]{#1}}

\title{Lightweight Interactive Proving inside an Automatic Program
  Verifier\thanks{Work partly supported by the Joint Laboratory
    ProofInUse (ANR-13-LAB3-0007,
    \protect\url{https://www.adacore.com/proofinuse}) of the French
    national research organization}}

\author{Sylvain Dailler \qquad\qquad Claude March\'e
\institute{Inria, Universit\'e Paris-Saclay, F-91120 Palaiseau\\
LRI, CNRS \& Univ.~Paris-Sud, F-91405 Orsay}
\and
Yannick Moy
\institute{AdaCore, F-75009 Paris}}

\def\titlerunning{Lightweight Interactive Proving inside an Automatic Program
  Verifier}
\def\authorrunning{S. Dailler, C. March\'e \& Y. Moy}

\begin{document}
\maketitle

\begin{abstract}

  Among formal methods, the deductive verification approach allows
  establishing the strongest possible formal guarantees on critical
  software. The downside is the cost in terms of human effort required
  to design adequate formal specifications and to successfully
  discharge the required proof obligations. To popularize deductive
  verification in an industrial software development environment, it is
  essential to provide means to progressively transition from simple
  and automated approaches to deductive verification. The SPARK
  environment, for development of critical software written in Ada,
  goes towards this goal by providing automated tools for formally
  proving that some code fulfills the requirements expressed in Ada
  contracts.

  In a program verifier that makes use of automatic provers to
  discharge the proof obligations, a need for some additional user
  interaction with proof tasks shows up: either to help analyzing the
  reason of a proof failure or, ultimately, to discharge the
  verification conditions that are out-of-reach of state-of-the-art
  automatic provers. Adding interactive proof features in SPARK
  appears to be complicated by the fact that the proof toolchain
  makes use of the independent, intermediate verification tool Why3,
  which is generic enough to accept multiple front-ends for different
  input languages. This paper reports on our approach to extend Why3
  with interactive proof features and also with a generic
  client-server infrastructure allowing integration of proof
  interaction into an external, front-end graphical user interface
  such as the one of SPARK.

\end{abstract}


\section{Introduction}

For the development of software with high safety and security
requirements, \emph{deductive program verification} is an approach
that provides access to the highest levels of guarantees. The
functional requirements are expressed using formal specification
languages. The conformance of the code with such specifications can be
established in a fairly automated setting using automatic program
verifiers available nowadays (such as Dafny, F$^\star$, KeY, KIV, OpenJML,
Verifast, Viper, Why3, etc.). Such verification tools typically
proceed by generating \emph{verification conditions} (VC for short):
mathematical formulas that need to be proven valid. Such VCs are
typically discharged using automated solvers, such as the SMT solvers
reasoning on \emph{Satisfiability Modulo Theories}.

A major issue preventing the diffusion of deductive verification in
industrial applications is the cost in terms of human effort required
to design adequate formal specifications and to successfully discharge
the VCs. To leverage this issue, it is important, when no SMT
solver is able to solve a given VC, to provide the user with means to
investigate the proof failure: is it because the code needs to be
fixed, is it because the program is not sufficiently annotated (e.g. a
missing loop invariant), or is it because the VC is too complex to be
discharged by automatic provers (e.g. an induction is needed). Among
the possible means is the generation of
counterexamples~\cite{hauzar16sefm}. Such a feature appears to be
useful in particular to fix trivial mistakes, but for complex
cases, its limitations quickly show up: because of the intrinsic
incompleteness of back-end solvers, or more pragmatically because
solvers proof search is typically interrupted after a given time limit
is reached, the counterexamples may be spurious or
absent~\cite{hauzar16sefm}.  Moreover, there is no easy mean to
distinguish a true bug from insufficiently detailed specifications (although
there is on-going research in that direction~\cite{petiot16tap}).


This paper presents an approach that we designed in the context of the
SPARK verifier for industrial development of safety-critical Ada
code~\cite{chapman14itp,mccormick15}. The goal is to provide
simplified interactions between the user and the failing VC, so as to
investigate a proof task without the need to rely on an external
interactive prover. A specificity of SPARK is that the underlying
toolchain from the given input Ada program to the VCs makes use of
the external intermediate language Why3~\cite{bobot14sttt} that itself
provides access to many different automated provers (mainly Alt-Ergo,
CVC4 and Z3) but also general purpose interactive theorem provers
(Coq, Isabelle/HOL, PVS). Indeed, an extreme mean to investigate a
proof failure is to launch an interactive theorem prover on the
failing VC and to start writing a manual proof. Such a process shows
up useful, mainly because writing the detailed steps, in which the user
believes the proof should work, typically helps to discover missing
elements in the specifications, and in such cases fixing the
annotations could finally help the SMT solvers to automatically
discharge the VC. Also, an ultimate situation is to finish the proof
completely using the underlying interactive
prover~\cite{berghofer12drops}. The main drawback in this process is
that users should be able to use the general-purpose back-end
interactive prover, forcing them to learn a completely different
environment, using its own syntax for formulas, and its specific proof
tactics to discharge proof tasks. Another drawback is that once the
user has switched to an external proof assistant to proceed with a
proof, then there is no easy mean to get back to the common
environment offered by Why3 to call other automatic provers on the
sub-goals generated by interactive proof tactics.

\subsection{Related Work}

The need for user interaction in the context of industrial use of
deductive verification is not new, this issue was identified and taken
into account early in the design of industrial tools. The KIV
environment, used in large realistic case studies in academia and
industry for more than 20 years, considered very early the importance
of combining automated and interactive proving~\cite{ahrendt98}. Other
early tools that provide some form of interactive proving are the
industrial tools, largely used in railway industry, Atelier
B~\cite{abrial03tphols} and Rodin~\cite{mehta07sefm}. The KeY
environment, somehow a successor of KIV, is designed to build proofs
interactively, with the possibility to call efficient automated
provers for solving some leaves of the
proof~\cite{hentschel16asea,hentschel16aseb}. In the context of
general purpose proof assistants, the need for adding automation to
their general interactive theorem proving process is evident, for
example in the environments ACL2, HOL Light, and more recently in
Isabelle/HOL where the so-called Sledgehammer subtool is able to finish
proofs using external SMT solvers~\cite{blanchette13jar}.

\subsection{Common Issues in Automated Program Verification}
\label{sec:usecases}

Interestingly, our analysis of the common situations where fully
automatic provers fail, and switching to proof interaction is needed,
is very similar to the analysis made in previous work mentioned above.
Here are the main identified cases:
\begin{itemize}
\item quantifier instantiation: a proof should be done by an adequate
  instantiation of a universally quantified hypothesis, that the
  automatic provers cannot discover. Providing the instantiation by
  hand helps.
  Similarly, for proving an existential goal, automated provers typically cannot
  guess the witness. This witness should be given by hand.
\item reasoning by cases: an explicit case reasoning can help the
  automatic provers.
\item controlling the context and the strategy of proof search: a prover would
  sometimes use most of its available time to try to solve the problem using a
  direction of thinking while a simple solution exists. Reducing the context
  manually can help.
\item inductive reasoning: some properties require reasoning by
  induction over an integer, an algebraic datatype, or on an
  inductively defined predicate; such reasoning steps are
  out-of-reach of common automated solvers, whereas applying an
  induction rule by hand usually results in sub-goals that can be
  automatically discharged.
\item non-linear integer arithmetic: it is typically hard for automatic provers,
  but a few manual proof steps can make such a proof
  feasible. Similarly, floating-point arithmetic is also very hard for automatic
  provers.
\end{itemize}

\subsection{Contributions and Overview of the Paper}

Our goals are shared with the above-mentioned related work. However,
in the context of SPARK, there is an additional issue that does
not show up in previous work. Indeed, in the previous work mentioned
above, the language of formulas and proof tasks (e.g. proof sequents)
is directly the language in which the user writes her input problem:
for example in the context of the B method, the logic language is B
set theory in which the code of the B machines is written; in
KeY, the underlying Dynamic Logic incorporates pieces of Java code to
verify. In the context of SPARK, we have the additional issue
that a proof task generated by the Why3 VC generator is written in
a language that is very different from the Ada input language.

Our approach proceeds in the following steps.
In Section~\ref{sec:why3}, we present what we added to the Why3
intermediate tool to provide interactive proving features.
Section~\ref{sec:spark} presents the use of interactive proof from
SPARK, inside GNAT Programming Studio, the Ada interface development environment.
Section~\ref{sec:conclusion} concludes and discusses remaining future work.





\section{Adding Interactive Proving in Why3}
\label{sec:why3}

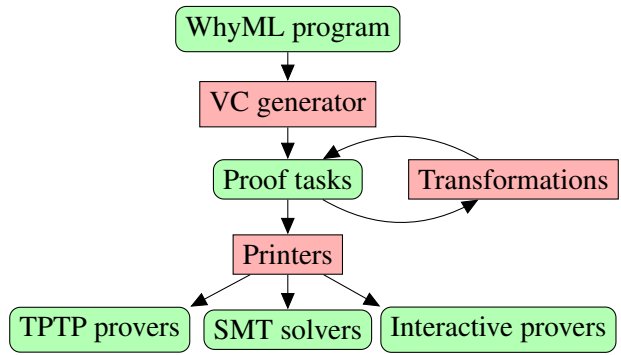
\begin{figure}[t]
  \centering
    \begin{tikzpicture}
    \node[bloc] at (5,5) (whyml) {WhyML program};
    \node[tool] at (5,4) (vcgen) {VC generator};
    \node[bloc] at (5,3) (tasks) {Proof tasks};
    \node[tool] at (8,3) (trans) {Transformations};
    \node[tool] at (5,2) (printers) {Printers};
    \node[bloc] at (5,1) (smt) {SMT solvers};
    \node[bloc] at (2.5,1) (tptp) {TPTP provers};
    \node[bloc] at (7.8,1) (itp) {Interactive provers};

    \draw[line] (whyml) -- (vcgen);
    \draw[line] (vcgen) -- (tasks);
    \draw[line] (tasks) to[in=210,out=-30] (trans);
    \draw[line] (trans) to[in=30,out=150] (tasks);
    \draw[line] (tasks) -- (printers);
    \draw[line] (printers) -- (smt);
    \draw[line] (printers) -- (tptp.north east);
    \draw[line] (printers) -- (itp.north west);
  \end{tikzpicture}
  \caption{Why3 general architecture}
  \label{fig:why3}
\end{figure}

Figure~\ref{fig:why3} presents a general overview of Why3's core
architecture. The input files contain code with formal specifications,
written in the dedicated language WhyML~\cite{filliatre13esop}, which
essentially consists of a set of functions or procedures annotated with
contracts (pre- and post-conditions, loop invariants, etc.). The VC
generator produces, from such a file, a set of \emph{proof tasks}. A
proof task, that we can denote as $\Gamma\vdash G$, consists in a set
$\Gamma$ of logical declarations of types, function symbols,
predicates, and hypotheses, and finally the logical formula $G$
representing the goal to prove. The soundness property of the VC
generator expresses that if all generated proof tasks are valid
logical statements, then the input program is safe: no runtime error
can arise and formal contracts are satisfied.

Consider the following toy example of a function written in WhyML.
\begin{lstlisting}[language=why3]
let f (a:array int) (x:int) : int
  requires { a.length >= 1000 }
  requires { 0 <= x <= 10 }
  requires { forall i. 0 <= 4*i+1 < a.length -> a[4*i+1] >= 0 }
  ensures { result >= 0 }
= let y = 2*x+1 in a[y*y]
\end{lstlisting}
The function \w{f} takes as parameters an array \w{a} and an integer
\w{x}. The first two pre-conditions express two simple requirements on
the size of array \w{a} and an interval of possible values for
\w{x}. The third pre-condition is a bit more complex, it expresses
that for indexes that are a multiple of 4 plus 1, the values stored in
\w{a} are non-negative. The function \w{f} returns an integer, denoted
as the keyword \w{result} in the post-condition, with a post-condition
expressing that
the value at the index returned is also non-negative. The code simply
returns the square of $2x+1$. For such a code, Why3 generates as the VC the formula
\begin{lstlisting}[language=why3]
forall a:array int, x:int.
 length a >= 1000 /\ (0 <= x /\ x <= 10) /\
 (forall i:int.
   0 <= ((4 * i) + 1) /\ ((4 * i) + 1) < length a -> a[(4 * i) + 1] >= 0) ->
   (let y = (2 * x) + 1 in
    let o = y * y in (0 <= o /\ o < length a) /\ a[o] >= 0)
\end{lstlisting}
As one may guess such a formula can quickly become unreadable when
code size grows, and is hardly suitable for human inspection.

\subsection{Proof Tasks and Transformations}

The core of Why3 comes as a software library, written in the OCaml
language and with a documented API, that proposes in particular
data-types for terms, formulas and proof tasks, and a large collection
of functions to operate on them. A central notion is the notion of
\emph{transformation}: an OCaml function that takes a proof task as
input and returns a set of proof tasks. All implemented
transformations are expected to be sound, in the sense that if all the
resulting proof tasks are valid, then the original task is valid
too. A simple example of such a transformation is \emph{splitting},
which basically transforms a task of the form
$\Gamma\vdash G_1\land G_2\land \cdots\land G_k$ into the set of tasks
$\Gamma\vdash G_i$ for $1\leq i\leq k$. The VC generator is designed
so as to produce a single proof task for each procedure or function of
the input code, as in the example above. To ease the visibility and
understanding of the resulting formula, a generalized splitting
transformation is typically applied, so as to decompose such a VC into
a set of simpler VCs for specific properties to check, \emph{e.g.}
checking if an array index is in bounds, checking the preservation of
a loop invariant, checking the pre-condition of a sub-procedure called,
etc.

\begin{figure}[t]
  \centering
\includegraphics[width=\textwidth]{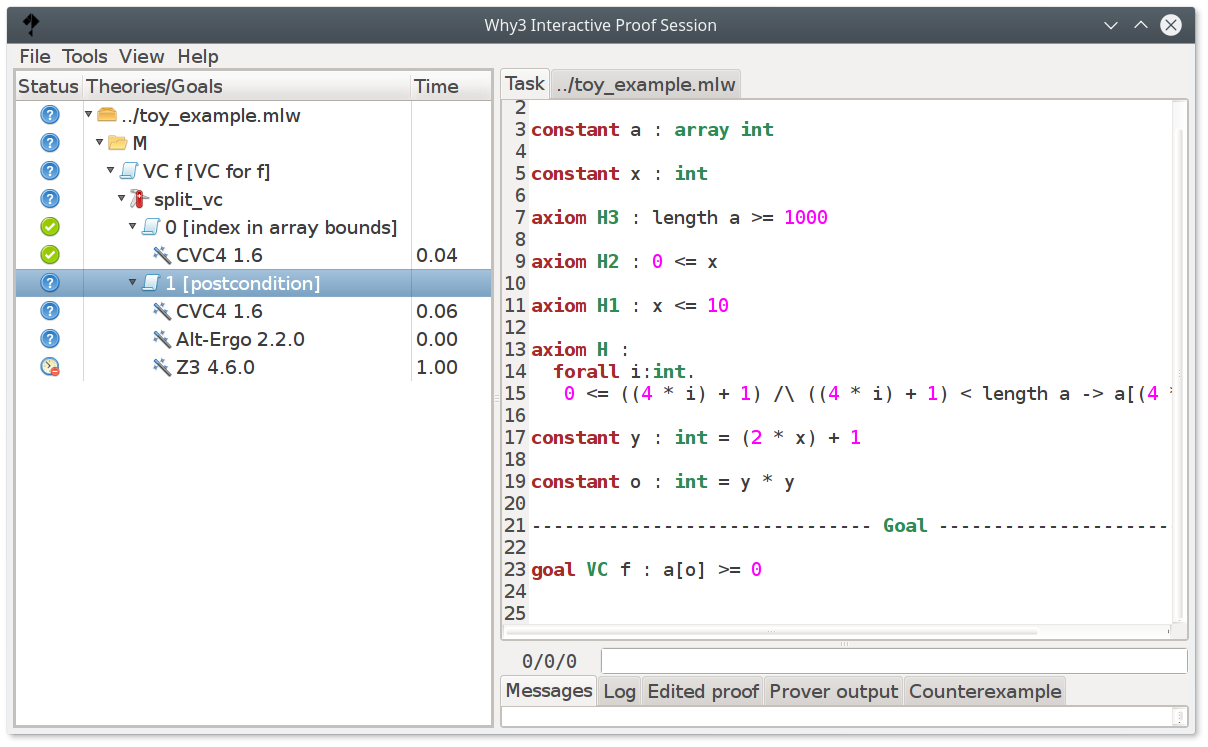}
\caption{Failed proof attempts  shown in Why3IDE}
\label{fig:ide1}
\end{figure}

Figure~\ref{fig:ide1} presents a screenshot of Why3's graphical
interface (Why3IDE) when given our toy program as input.  The left part of that
window is the \emph{proof task tree}. The current selected line is for
the proof task after applying the transformation named
\w{split_vc} which implements the generalized splitting
transformation mentioned above.

Indeed the role of transformations is two-fold. The first role is to
simplify the given task (such as the splitting above). In such
a case a transformation is applied on user's request inside the
graphical interface. The second role is to preprocess a task before
sending it to an external prover: typically an external prover does
not support all features of Why3's logic, such as polymorphic types,
algebraic data types and the corresponding pattern-matching, recursive
or inductive definition of predicates, etc. Hence, when the user wants
to invoke an external prover on a given task, Why3 transparently
applies some transformations to make the proof task fit into the logic
of the target prover, before using an appropriate \emph{printer}
suitable for the back-end prover input syntax (e.g. SMT-LIB).  On
Figure~\ref{fig:ide1} it can be seen that the provers Alt-Ergo, CVC4
and Z3 were invoked but none of them were able to discharge the
expected post-condition. It is likely that the combination of
non-linear arithmetic and the need for finding an appropriate
instantiation for the hypothesis \w{H} is the reason why they fail.
Mixing quantifiers and arithmetic makes very difficult goals to prove
for all categories of automatic provers: SMT solvers quantifier
handling is based on triggers, that do not interact well with
arithmetic, while TPTP provers have a more powerful handling of
quantifiers but do not support arithmetic. This corresponds to the
limitation of automatic provers that we called \emph{quantifier instantiation}
in Section~\ref{sec:usecases}.

On top of the core architecture of Figure~\ref{fig:why3}, Why3
features two major components: the \emph{proof session} manager and
the graphical interface.  Adding support for interactive proving in
this global architecture requires extensions that we detail in the
subsections below: extensions of the GUI, extension of the
transformation setting of the core architecture, and extensions in the
proof session manager.
A feature, that has important consequences on our approach presented
below for adding interactive proving in Why3, is the genericity of
transformations handling. The Why3 library is designed so that an
additional user-written transformation can be dynamically loaded at
run-time, using a mechanism of registration with a name.

\subsection{Extending User Interface}
\label{sec:interface}

\begin{figure}[t]
  \centering
\includegraphics[width=\textwidth]{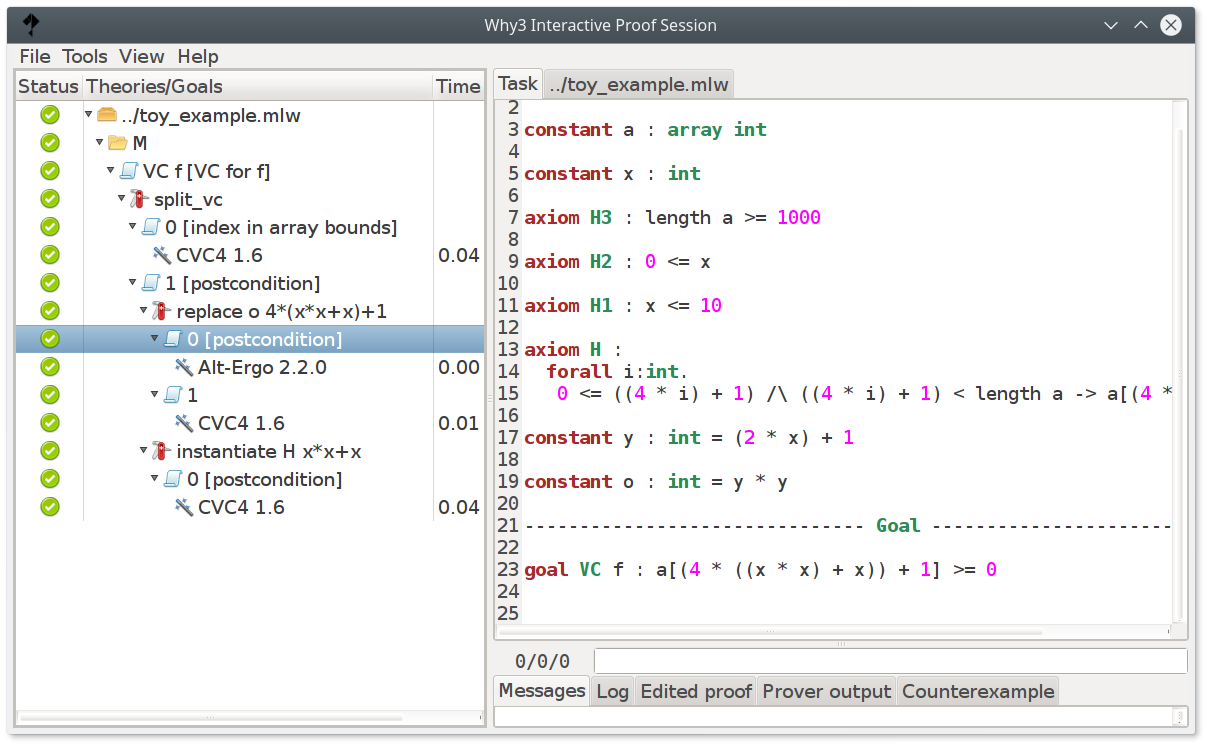}
\caption{Completed proof using two possible transformations with arguments}
\label{fig:ide2}
\end{figure}

As shown in Figure~\ref{fig:ide1}, the graphical interface is
naturally where proof tasks are displayed and where the user can
decide which transformations to apply and which prover to call. Below
the top-right part of the window, where the current proof task is
displayed, and above the bottom-right part where different kinds of
messages are displayed, we added a kind of command-line input field
where the user can input arbitrary character text to form a command.
If we consider our toy example, one possibility to progress towards
proving the goal is to replace \w{y*y} by the term \w{4*(x*x+x)+1}. To
achieve that, the user can directly input the text
\begin{lstlisting}[language=why3]
replace y*y 4*(x*x+x)+1
\end{lstlisting}
in the input field and hit the return key. An alternative possible
transformation would be to instantiate the hypothesis \w{H} with
\w{x*x+x}, which can be done with the input
\begin{lstlisting}[language=why3]
instantiate H x*x+x
\end{lstlisting}
Figure~\ref{fig:ide2} displays the GUI after trying both
transformations. As seen on left, the transformation \w{replace y*y
  4*(x*x+x)+1} generated two subgoals: first proving the formula after
the replacement, second proving that \w{y*y=4*(x*x+x)+1}. Both
subgoals are proved by Alt-Ergo. Similarly, the transformation
``\w{instantiate H x*x+x}'' generates one subgoal where an additional
hypothesis is present (the instance of \w{H} for the given particular
value for \w{i}) and is also proved by Alt-Ergo.
As can be seen, applying any of these two transformations is enough to
finish the proof automatically.

Even if this mechanism using a textual interface may seem
old-fashioned, it permits a lot of genericity. We'll see below how it
simplifies the communication with a front-end such as SPARK.  It also
permits a lot of extra features that show themselves important in
practice, such as searching in the proof context.

\subsection{Adding Parameters to Proof Transformations}
\label{sec:transformations}

A central design choice towards the introduction of interactive proofs
is to reuse the existing infrastructure of transformations. Basically,
since that infrastructure already allows the user to select among a
given set of transformations to apply on a given proof task, we just
have to extend this set, after extending the concept of
transformations so that they can take parameters, like the two
transformations used above: \w{replace} is a transformation that takes
two terms as input. \w{instantiate} takes a hypothesis name and a
term. We faced two main issues for this extension.  First, the
transformations can take various objects as parameters: a term, a
formula, a name, a string, etc. It means that at the level of the API,
we need a typing mechanism in order to declare what are the right
types of objects to pass as parameters. Second, at the level of the
interface, the data submitted by the user are just textual, so we need
a generic infrastructure to parse them and turn them into objects of
the right kind.

At the level of the API, in order to handle the large variability of
the kinds of transformation parameters, we were able to use the
advanced concept of GADT (Generalized Abstract Data Types). An excerpt
of the new declaration of transformation type in the API is as follows (the
real one has 20 constructors):
\begin{lstlisting}[language=Caml]
type _ trans_typ =
  | Ttrans_l    : (task -> task list) trans_typ
  (** transformation with no argument, and many resulting tasks *)
  | Tstring     : 'a trans_typ -> (string -> 'a) trans_typ
  (** transformation with a string as argument *)
  | Tprsymbol   : 'a trans_typ -> (Decl.prsymbol -> 'a) trans_typ
  (** transformation with a Why3 proposition symbol as argument *)
  | Tterm       : 'a trans_typ -> (Term.term -> 'a) trans_typ
  (** transformation with a Why3 term as argument *)
  | Topt        : string * ('a -> 'c) trans_typ -> ('a option -> 'c) trans_typ
  (** transformation with an optional argument. The first string is
      the keyword introducing that optional argument*)
\end{lstlisting}
To implement a transformation like \verb|instantiate| above, we first have to
program it with an OCaml function, say \w{inst}, of type \w{prsymbol -> term -> task
  list}, and then register it under the proper name as follows
\begin{lstlisting}[language=Caml]
wrap_and_register "instantiate" (Tprsymbol (Tterm Ttrans_l)) inst
\end{lstlisting}
Not only will it make the transformation available for use in the
interface, but it will automatically proceed with the parsing, name
resolution and typing of the textual arguments, as given by the type
\w{(Tprsymbol (Tterm Ttrans_l))}. This mechanism based on GADTs is
powerful enough to handle optional parameters. For example, the
\w{replace} transformation is declared with type
\begin{lstlisting}[language=Caml]
(Tterm (Tterm (Topt ("in", Tprsymbol Ttrans_l))))
\end{lstlisting}
which means that a third optional argument is allowed, of type
\w{prsymbol} and introduced by the keyword \verb|in| to say that the
replacement should be done in the hypothesis of the given name instead
of the goal, e.g. ``\verb|replace (length a) 1000 in H|''.  Notice the
large genericity of this mechanism, in particular the keyword used for
introducing the optional argument. The genericity also comes from the
\w{wrap_and_register} function which is defined once and for all and
does all the hard job of parsing and typing arguments. In particular,
the resolution of variable names given as arguments had to be
carefully made consistent with the printing of the task, which can
rename variables.

Here is a quick summary of the major transformations with parameters
that we added in Why3. They are supposed to cover the major
needs for interaction as already listed in Section~\ref{sec:usecases}.
\begin{itemize}
\item case analysis on a formula (\verb|case|~$P$), on algebraic data
  (\verb|destruct_alg|~$t$), decomposition on propositional structure of a hypothesis (\verb|destruct|~$H$). 
For example, the transformation ``$\texttt{case}~P$'' where $P$ is an arbitrary formula would be
transforming a task $\Gamma\vdash G$ into the two tasks
$\Gamma,P\vdash G$ and $\Gamma,\neg P\vdash G$.
\item introduction of an auxiliary hypothesis (\verb|cut|~$P$, \verb|assert|~$P$)
  or term (\verb|pose|~$x~t$)
\item induction on integers, on inductive predicates
\item instantiation as seen above (\verb|instantiate|~$H$~$t_1,\ldots,t_k$), including existential case (\verb|exists|~$t$), or via direct application of a hypothesis to a goal (\verb|apply|~$H$)
\item various rewriting and computation transformations: \verb|rewrite|~$H$~(\verb|in| $H'$),
  \verb|replace|~$t_1~t_2$~(\verb|in|~$H$), \verb|subst|~$x$, \verb|subst_all|, etc.
\item context handling: \verb|remove|~$H_1,\ldots,H_k$, \verb|clear_but ...|
\item unfolding a definition: \verb|unfold|~$f$
\item import an extra theory: \verb|use_th|~$T$
\end{itemize}

\subsection{Extending The Proof Session Mechanism}

A proof session is essentially a record of all the
proof tasks generated from a given input file, and also a record of
all transformations applied to these tasks. It is indeed an internal
representation of the proof task tree displayed on the left part of
Figure~\ref{fig:ide1}. Such a session can be stored on disk, and can
be reloaded to a former state by the user. A crucial feature of the
session manager is to manage the changes if the input file is modified
(e.g. more annotations are added): the manager implements a clever
and sound \emph{merging} operation to discover which parts of the proof session
can be reused, which tasks are modified, and which external proofs should
be replayed~\cite{bobot13vstte}.

The Why3 session files do not store any internal representation to
avoid any problem when the Why3 tools themselves evolve. Accordingly,
we decided that the arguments of transformations should be stored
under their textual form too. This definitely avoids potential
problems with changes in internal representations, but still some
problems with renaming could occur. For example, an automatically
introduced name for any hypothesis, say \w{H1}, could perfectly be
renamed into \w{H2}, e.g. if an extra annotation is added in the
source code. It is thus perfectly normal that from time to time, while
reloading a proof session, a transformation with argument does not
apply anymore. In order to avoid the loss of any sub-proof tree, we
implemented the new notion of \emph{detached nodes} in the proof task
tree. These nodes are a record of the state of the previous session,
but without any corresponding proof task. We then implemented a
mechanism to copy and paste fragments of proof trees from one node to
another. This copy-paste mechanism showed itself very useful in
practice for maintaining interactive proofs.

\subsection{Examples}
\label{sec:why3examples}

We evaluated the new interactive proof features of Why3 on prior
examples where some VCs could not be discharged except using the Coq
proof assistant.  Bobot et al. paper~\cite{bobot14sttt} illustrated
the use of Why3 on the three challenges of the VerifyThis competition
in 2012. On each of these case studies, at least one Coq proof was
required. We have reconsidered the first challenge (Longest Repeated
Substring) and were able to replace all three Coq proofs with
interactive proofs. Interestingly, only very few transformations were
needed, because we quickly arrived at simpler subgoals that are
discharged by automatic solvers.

Another illustrative example is the proof of Dijkstra's shortest path
algorithm on graphs: again, we were able to replace Coq proofs with
interactive transformations and automatic solvers. We noticed that not
only does it simplify the proofs, they are now easier to maintain in
case of a change in Why3 implementation or standard library.

We still have to evaluate to what extent the interaction using
transformations with argument is easy to use for regular users.
Moreover, we need more practice in order to see if this mechanism is
of effective help for debugging proofs, as explained in the introduction.

\section{Adding Interactive Proving in SPARK}
\label{sec:spark}

Although the SPARK verifier, called GNATprove, is based on Why3 for generating
VCs and proving them, there is a large gap between the SPARK and WhyML
programming languages. Therefore, the interactive proof features of Why3 cannot
be used directly by SPARK users. The same issue arose in the past with the
counterexample generation features of Why3, which required translation back to
SPARK for use inside GNATprove~\cite{hauzar16sefm}. That issue also involved
interactions with users inside different IDEs, Why3IDE for Why3 users and GNAT
Programming Studio for SPARK users, but that interaction was one-way
only: the counterexamples output by provers were translated back to either Why3
or SPARK syntax and displayed in their respective IDE. Here, we need a two-way
interaction where users can input commands (possibly with elements of the code
as parameters) and Why3 returns a modified set of proof tasks.

We start by presenting a simple SPARK program that cannot be proved
with automatic provers in Section~\ref{subsec:example}. Then we describe in
Section~\ref{subsec:client-server} the client-server architecture that we have
adopted for two-way communication between the IDE for SPARK programs and
Why3. We detail in Section~\ref{subsec:printing-parsing} how we translate back
and forth between user-level names in SPARK and internal names in
Why3. Finally, we explain in Section~\ref{subsec:application-example} how to
complete the proof of our example.

\subsection{Unprovable Code Example}
\label{subsec:example}

The code in Figure~\ref{fig:unproved} is a simplified version of an excerpt from
a bounded string library. This code contains a post-condition that cannot be
proved today by any prover available with SPARK (Alt-Ergo, CVC4, Z3).
The
reason for this unproved property is characteristic of the kind of problems
faced by users of a technology like SPARK. It is in the class of problems called \emph{quantifier instantiation} already presented in Section~\ref{sec:why3}.

The code in Figure~\ref{fig:unproved} computes the location of a
sub-list of integers within a list of integers, when the sub-list is
contained in the list. Lists are implemented here as SPARK arrays
starting at index 1 and ranging over positive indexes. The
post-condition of function \wa{Location} introduced by \wa{Post}
states the following properties: the result of the function ranges
from 0 to the length of the list; a positive result is used when the
sub-list is contained in the list, and value 0 is used as result
otherwise. For the sake of simplicity, we do not state the complete
post-condition of \wa{Location} (that would make precise that the
resulting index is the location of the match), but this could be done
easily.

This post-condition relies on the definition in function \wa{Contains} of what
it means for a sub-list \wa{Fragment} to be contained in a list
\wa{Within}. This function is only used in specifications, which is enforced by
marking it as \wa{Ghost}. It is defined as an expression function, quantifying
with \verb|for some| (Ada existential quantification) over a range of scalar values the property that the sub-list
\wa{Fragment} is equal to a moving slice of the list \wa{Within (K .. (K - 1 +
  Fragment'Length))}. This last expression is the slice of array \wa{Within}
from index \wa{K} to index \wa{(K - 1 + Fragment'Length)}.

For the sake of simplicity, function \wa{Location} naively iterates over each
possible location for a match and tests for equality of the corresponding slice
with the argument sub-list. The loop invariant repeats the post-condition and
specializes it for the \wa{K}$^{th}$ iteration of the loop, so that the loop
invariant is itself provable and can be used to prove the post-condition.

\begin{figure}[t]
\begin{lstlisting}[label={lst:array}, style=myAda]
   type List is array (Positive range <>) of Integer
     with Predicate => List'First = 1;
   subtype Natural_Index is Integer range 0 .. Positive'Last;

   function Contains (Within : List; Fragment : List) return Boolean is
     (Fragment'Length in 1 .. Within'Length and then
        (for some K in 1 .. (Within'Length - Fragment'Length + 1) =>
           Within (K .. (K - 1 + Fragment'Length)) = Fragment))
   with Ghost;

   function Location (Fragment : List; Within : List) return Natural_Index
   with
     Post => Location'Result in 0 .. Within'Length and then
       (if Contains (Within, Fragment) then
          Location'Result > 0
        else
          Location'Result = 0)
   is begin
      if Fragment'Length in 1 .. Within'Length then
         for K in 1 .. (Within'Length - Fragment'Length + 1) loop
            if Within (K .. (K - 1 + Fragment'Length)) = Fragment then
               return K;
            end if;
            pragma Loop_Invariant
              (for all J in 1 .. K =>
                 Within (J .. (J - 1 + Fragment'Length)) /= Fragment);
         end loop;
      end if;
      return 0;
   end Location;
\end{lstlisting}
\caption{Example of SPARK code with an unprovable post-condition}
\label{fig:unproved}
\end{figure}

When running GNATprove on this code, it proves automatically the absence of
runtime errors (no integer overflows, no array access out of bounds, no other
runtime check failures), as well as the loop invariant in function
\wa{Location}, but it does not prove the post-condition of that function.

The user interaction is simple here. The user requests that this
program is verified by GNATprove inside GPS (GNAT Programming Studio),
and a few seconds later receives the output of the tool as messages
attached to program lines. Between these two instants, GPS called
GNATprove; GNATprove translated the SPARK program into an equivalent
WhyML program w.r.t. axiomatic semantics for generation of VCs; an
internal program using Why3 API successively generated VCs by calling
Why3 VC generator and
dispatched VCs to provers;
this internal program collected
the output of provers and returned the overall results to GNATprove; GNATprove
generated and adapted results for GPS, that displayed these results to
the user.

All this work occurred transparently for the user, who never had to see the
generated WhyML code in Why3IDE, or to launch Why3 commands in a terminal. A
less-than-ideal process for completing the proof of the post-condition would
consist in asking the user to open the session file generated by Why3 in
Why3IDE to complete the proof using the interactive proof feature described in
Section~\ref{sec:why3}.  While this would work, this is not a suitable solution in
an industrial context. Indeed, asking users to interact directly with a
generated artifact in a different language (the Why3 file) is akin to asking
them to debug their programs at assembly level. While this is possible and
sometimes useful, it is best left to rare occasions when this is really needed,
and instead interaction should be done as much as possible at source code level.

\subsection{Client-Server Architecture}
\label{subsec:client-server}

\tikzstyle{arrowright} = [draw, line width=1.5pt, color=red, ->]
\tikzstyle{arrowleft} = [draw, line width=1.5pt, color=blue, <-]
\tikzstyle{doublearrow} = [draw, line width=1.5pt, <->]

\begin{wrapfigure}{R}{0.4\textwidth}
  \centering
  \scalebox{.9}{
  \def\distnode{1cm}
  \def\distdouble{2cm}

  \begin{tikzpicture}[node distance=2.5cm,auto,>=latex']
    \node (ul) {GPS IDE};
    \node[right of=ul, node distance=5cm] (ur) {\hspace*{-20mm}Why3 interactive server};
    \node[below of=ul, node distance=10cm] (bl) {time};
    \node[right of=bl, node distance=5cm] (br) {time};
    \node[below right of=ul, node distance=1cm] (l1) {};
    \node[below of=l1, node distance=0.5cm] (l2) {};
    \node[below of=l2, node distance=\distnode] (l3) {};
    \node[below of=l3, node distance=\distnode] (l4) {};
    \node[below of=l4, node distance=\distnode] (l5) {};
    \node[below of=l5, node distance=\distdouble] (l6) {};
    \node[below of=l6, node distance=\distdouble] (l7) {};
    \node[below of=l7, node distance=\distnode] (l8) {};
    \node[below left of=ur, node distance=1cm] (r1) {};
    \node[below of=r1, node distance=0.5cm] (r2) {};
    \node[below of=r2, node distance=\distnode] (r3) {};
    \node[below of=r3, node distance=\distnode] (r4) {};
    \node[below of=r4, node distance=\distnode] (r5) {};
    \node[below of=r5, node distance=\distdouble] (r6) {};
    \node[below of=r6, node distance=\distdouble] (r7) {};
    \node[below of=r7, node distance=\distnode] (r8) {};

    \draw[arrowright] (l2) edge node {Launch the server} (r2);
    \draw[arrowleft] (l3) edge node {Notify starting proof tree} (r3);
    \draw[arrowright] (l4) edge node {Ask for proof task $i$} (r4);
    \draw[arrowleft] (l5) edge node {Send the task} (r5);
    \draw[doublearrow] (l6) edge node {...} (r6);

    \draw[arrowright] (l7) edge node {Exit request} (r7);
    \path[arrowleft] (l8) edge node {Save session and stops} (r8);

    \path[->] (ul) edge node {} (bl);
    \path[->] (ur) edge node {} (br);

  \end{tikzpicture}
}
\caption{Schematic of the interactions between IDE and server}
\label{fig:client_server}
\end{wrapfigure}
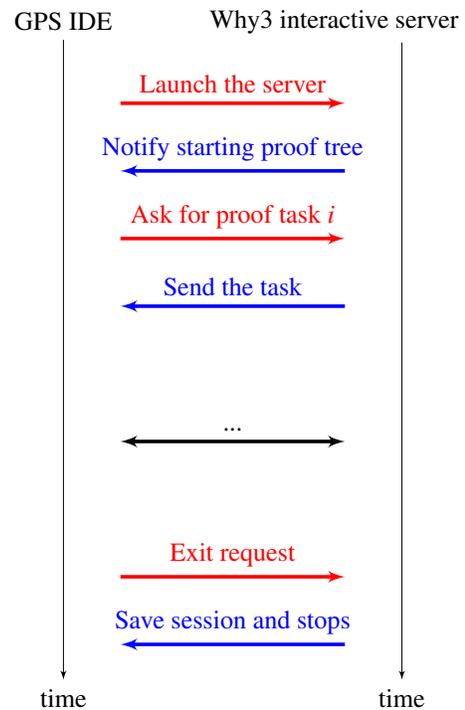

Most modern IDEs like GPS provide client-server interfaces with various tools such
as debuggers or, in the context of formal verification, with proof assistants
such as Emacs ProofGeneral support for Coq, Isabelle, Lego, HOL, etc. We have
adopted a client-server architecture to allow two-way communications between
the interactive proof module (acting as a server) and the client IDE, which can
be Why3IDE or GPS here. The server handles requests from the user (through the
IDE), such as proof transformations (see Section~\ref{sec:transformations}) and direct calls
to provers. After the requested action terminates, the server informs the IDE
of changes to the proof task tree.

For the integration in GPS, we developed both a wrapper in OCaml for the
underlying service to act as the server, and a plugin in Python to communicate
with the server from GPS. The server takes the session file as initial
argument, gets its input in JSON format on standard input, calls Why3 core
services, updates the session file accordingly, and returns its output in JSON
format on the standard output. The plugin modifies GPS interface to add a
console window for command-line interaction, a window to display VCs and a
window to display the proof task tree.
The plugin translates requests made by the user on the command-line
interface into JSON requests that are sent to the server, and translates back
the server notifications into updates of the graphical user interface (adding
nodes in the proof task tree, changing the VC, etc.).
For example, as seen in Figure~\ref{fig:client_server}, GPS first starts the
server, then the server returns the initial proof task tree which is printed
by GPS. When the user clicks a node, GPS asks for the corresponding proof task
which the server returns and GPS then prints it for the user. This goes on
until the user exits manual proof which GPS interprets by sending the exit
request: Why3 interactive server ends its execution. The death of the process
is detected by GPS which goes back to its normal interface. A schematic of the
interactions between the IDE and the server is depicted in
Figure~\ref{fig:client_server}.
With very little effort, we transformed a generic IDE such as GPS into an
elementary proof assistant.


\begin{figure}[t]
  \centering
\includegraphics[width=1.0\textwidth]{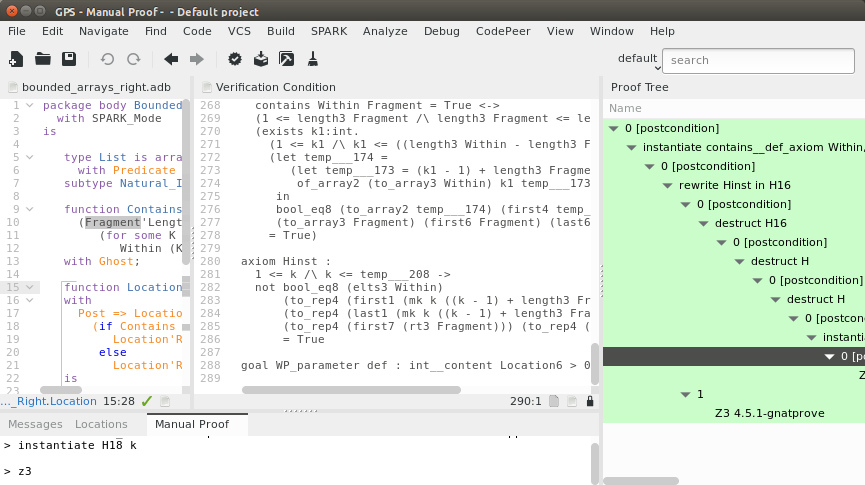}
\caption{Example of interactive proof in GPS}
\label{fig:gps_proved}
\end{figure}

As seen on Figure~\ref{fig:gps_proved}, the user interface
in GPS is similar to the one in Why3IDE presented in Figure~\ref{fig:ide1}. The
same windows are present but they are not located at the same place. From left
to right, we can see the SPARK code window, the proof task window and the proof
task tree window. The command-line console is displayed as a bottom panel
sharing its window with other panels for Messages (tool output) and Locations
(tool messages). The user can type commands for applying transformations and
calling provers inside the command-line console, similar to what we saw for
Why3IDE.

One can observe that the VC from Figure~\ref{fig:gps_proved} is much more
complex than the simple VC we generated with Why3 in
Figure~\ref{fig:ide1}. This is mainly a consequence of the complexity of the
WhyML code generated from SPARK. Simple data structures and control flow in
SPARK are modeled with much more complex data structures and control flow in
WhyML to encode the semantic features of SPARK. This complexity gets
exposed in the VC generated by Why3 VC generator, as it combines the complexity of
data structures and control flow. This inherent complexity cannot be eliminated
and must be dealt with to present SPARK users with understandable VCs.

\subsection{Printing of Proof Tasks and Parsing of Transformation Arguments}
\label{subsec:printing-parsing}

In order for users to be able to relate the proof task to their code, it is
necessary to use the names of source SPARK entities in the proof task instead
of the generated Why3 names. This is achieved by creating a mapping from Why3
names to their source SPARK name during the generation of Why3 code from
SPARK. This mapping is embedded in the WhyML code using \emph{labels}, a generic
mechanism in Why3 to attach strings to terms. For example, in the following
code snippet the label \w{"name:Not_y"} is attached to the identifier~\w{y}:
\begin{lstlisting}[language=why3]
let f (a:array int) (x:int) : int
  ensures {result = 20}
= let y "name:Not_y" = 2*x+1 in y*y
\end{lstlisting}

Labels on terms are preserved by VC generation and transformations. For
example, the label \w{"name:Not_y"} remains attached to the constants derived
from \w{y} in the final VC obtained after VC generation and
transformations. Thus, when printing a proof task, the name of an identifier can be
replaced by the name found in the attached label when there is one. Thus we get
the following proof task where the source SPARK name \wa{Not_y} is used instead
of the generated Why3 name \wa{y}:

\begin{lstlisting}[language=why3]
constant x : int
constant Not_y : int = (2 * x) + 1
goal VC f : (Not_y * Not_y) = 20
\end{lstlisting}

Different Why3 names with the same name label are currently distinguished by
appending a unique number to the source SPARK name.
The consecutive problem of interpreting the names of SPARK entities as Why3
names occurs when the user types a command with arguments referring to the
names of SPARK entities. Why3 uses here the inverse map from distinguished
SPARK names to Why3 names associated to a given proof task to translate
automatically the arguments of transformations.

\subsection{Application to the Unprovable Code Example}
\label{subsec:application-example}

With the interactive proof interface in GPS, we complete the proof of the
post-condition of \wa{Location} from Section~\ref{subsec:example}.
For the sake of simplicity, we only describe the proof part of the
\verb|then| branch in this subsection, shown in the
screenshot in Figure~\ref{fig:gps_proved}.
The root of the proof task tree is green,
showing that the initial VC was fully proved.

The interactive proof proceeds by deriving a contradiction from the loop
invariant property in the last iteration of the loop and the condition of the
\wa{then} branch \wa{Contains (Within, Fragment)}. This requires unfolding the
definition of \wa{Contains} and finding suitable instances of quantified
properties. In this case, the proof task tree is linear and a complete proof
script is the following:

\begin{verbatim}
instantiate contains__def_axiom Within,Fragment
rewrite Hinst in H16
destruct H16
destruct H
destruct H
instantiate H18 k
\end{verbatim}

The intuition of the proof is that we need to explain to the tool how to combine
the hypotheses coming from the loop invariant and the definition of \wa{Contains}.
The first two transformations (\verb|instantiate|,
\verb|rewrite|) are used to make the axiomatic definition of \wa{Contains},
applied to the right arguments, appear as a hypothesis in the context.  The
three calls to the \verb|destruct| transformation are used to destruct the head connective of
the hypothesis that appeared. They transform $\Gamma, H: (A \wedge \exists
k. P(k)) \vdash G$
into $\Gamma, H: A, k: \verb|int|, H_1: P(k) \vdash G$.
The objective is to use $k$ as an instance for the loop invariant
property (to make the contradiction appear). This is exactly what transformation
``\wa{instantiate H18 k}'' does.
The quantification disappears and SMT solvers can now solve
the goal.
This completes the proof of the \verb|then| branch of the
post-condition of \wa{Location}. The \verb|else| branch is handled
similarly, which completes the proof of the program.

\section{Conclusions and Future Work}
\label{sec:conclusion}

We brought interactive proving features to the SPARK verification
environment for Ada. This was done by conservatively extending the
intermediate Why3 tool by allowing to pass arguments to proof task
transformations. The design is generic enough to allow simple addition
of new transformations via Why3's API. The proof session mechanism and
the graphical interface have been extended to allow simple user
interaction and facilities for proof maintenance. A few program
examples were re-proved, showing that former external interactive
proofs using Coq can be substituted with light interactive proofs
in our new setting. The user interface of Why3 was redesigned under
the form of a generic client-server architecture, allowing to bring
interactive proving features to the SPARK front-end inside the regular
GPS graphical interface.




Future Work will go into several directions. First, we certainly want to
enlarge the set of transformations with arguments, to cover more needs
for interactive proofs. The needed transformations will be identified
from practice. Notice that others are already reusing our API to
implement new transformations, for example for doing proofs by
reflection for complex non-linear goals~\cite{melquiond18ijcar}. Second, we plan to reuse the
client-server architecture to provide an alternative interface for
Why3, this time within a web browser. We also plan to bring
the interactive proof feature to the Frama-C front-end for C code, by
augmenting the existing Jessie plug-in of Frama-C that uses Why3
internally.  A third longer-term work is to allow for more
customization of the printing of tasks and the parsing of
transformation arguments: from SPARK, we would like the terms in proof
tasks expressed in a more Ada-like syntax, in particular for
arrays. For this we will need to design in Why3's generic API a
possibility to register printers and parsers.  A fourth even
longer-term issue is the question of trust in the implemented
transformations. Since we implement more and more complex OCaml code
for that purpose, a general need for verifying the soundness of the
transformations shows up. It is likely that we will need to design a
language of proof terms or proof certificates to achieve this
long-term goal.

\bibliographystyle{eptcs}
\bibliography{generated}

\end{document}